\documentclass[reprint,showpacs,showkeys,preprintnumbers,amsmath,amssymb,aip,apl]{revtex4-1}

\usepackage{graphicx}
\usepackage[english]{babel}
\usepackage{xspace}
\usepackage[normalem]{ulem}
\usepackage{color}
\usepackage{float}
\usepackage{amsmath}
\usepackage{footnote}

\begin{document}

\title{ Investigation  of the Mn$_{3-\delta}$Ga/MgO interface for  magnetic tunneling junctions }

\author {C. E. ViolBarbosa  }
\email{carlos.barbosa@cpfs.mpg.de}
\affiliation {Max-Planck-Institut f\"{u}r
Chemische Physik fester Stoffe, N\"{o}thnitzer Strasse 40, 01187 Dresden, Germany}

\author { S. Ouardi \footnote{C.E. ViolBarbosa and S. Ouardi contributed equally to this work.}}
\affiliation {Max-Planck-Institut f\"{u}r
Chemische Physik fester Stoffe, N\"{o}thnitzer Strasse 40, 01187 Dresden, Germany}

\author{T. Kubota}
\affiliation{WPI Advanced Institute for Materials Research Tohoku University, 980-8577,  Sendai, Japan}

\author {S. Mizukami}
\affiliation{WPI Advanced Institute for Materials Research Tohoku University, 980-8577,  Sendai, Japan}
%Shigemi

\author {G. H. Fecher}
\affiliation {Max-Planck-Institut f\"{u}r
Chemische Physik fester Stoffe, N\"{o}thnitzer Strasse 40, 01187 Dresden, Germany}

\author{T.  Miyazaki}
\affiliation{WPI Advanced Institute for Materials Research Tohoku University, 980-8577,  Sendai, Japan}

\author{X. Kozina}
\affiliation{JASRI,SPring-8, Sayo-cho, Hyogo 679-5198, Japan}

\author{E. Ikenaga}
\affiliation{JASRI,SPring-8, Sayo-cho, Hyogo 679-5198, Japan}

\author {C. Felser}
\affiliation {Max-Planck-Institut f\"{u}r
Chemische Physik fester Stoffe, N\"{o}thnitzer Strasse 40,01187 Dresden, Germany}

%\date{received 15$^{th}$ July 2013}
\begin{abstract}

The Mn$_3$Ga Heusler compound and related alloys are the most promising materials for the realization of spin-transfer-torque switching
in magneto tunneling junctions. Improved performance can be achieved by high quality interfaces in these multilayered structured devices.
  In this context, the interface between Mn$_{1.63}$Ga and MgO is of particular interest because of its spin polarization properties in tunneling junctions.
We performed a chemical characterization of the MgO/Mn$_{1.63}$Ga junction by hard x-ray photoelectron spectroscopy.
The experiment
indicated the formation of Ga-O bonds at the interface and evidenced changes in the local environment of  Mn atoms in the proximity
of the MgO film. In addition, we show that the insertion of a metallic Mg-layer interfacing the MgO and Mn--Ga film strongly suppresses the
oxidation of gallium.

\end{abstract}

 \maketitle

\maketitle

The engineering of a high performance magnetic tunneling junction (MTJ) is crucial for the emergence of a new class of non-volatile memories and spintronic devices.
 Directly related with these recent technologies, Mn$_2$YZ Heusler compounds,
where Y is a transition metal and Z is a main group element,  have shown interesting properties such as high spin polarization, high
Curie temperature (Tc), low net magnetic moment, and strong magneto-crystalline anisotropy\cite{Graf:2011}.
Mn$_3$Ga and related alloys appear as  promising materials in the realization of  switching type spin-transfer-torque mangnetoresistive random access memories (STT-MRAMs).  Mn--Ga films present perpendicular magneto anisotropy (PMA) and a nearly compensated ferrimagnetic phase\cite{djayaprawira:2005,Ikeda:2010,Kubota1:2011,Mizukami:2012,Bai:2012}. These  characteristics are important for the reduction of the switching  current
  and downscaling of the dimensions while maintaining thermal stability in  STT-MRAM devices\cite{Balke:2007}.
The full understanding and control of growth properties of the Mn--Ga/MgO system will open up avenues for the exploitation of Heusler compounds in ultra-high density memory technologies.

The performance of MTJ devices is strongly influenced by the quality of the film interfaces. Long-range ordering  of the two-dimensional interfaces is required for conservation of the electron moment transverse to the tunneling current propagation ($k_\|$).\cite{Zhang:2004} The integration of Heusler compounds in MTJs has been successfully demonstrated for
 Co$_2$MnSi\cite{Krumme:2012,Tsunegi:2008}. The great advantage of the Mn$_2$YZ family over other Heuslers is the
 tetragonal distortion of the crystalline structure leading to the PMA. This allows an unique combination of high-spin polarization with
 perpendicular magnetization in MTJs.
 Recent studies on the engineering of MTJs using Mn--Ga
alloys by Kubota \emph{et al.} \cite{Kubota2:2011, Kubota3:2011, Kubota1:2011} resulted in the optimized layered structure:
Cr buffer/Mn--Ga/MgO/CoFe. The MgO  grows compressed on the Mn--Ga
structure with a lattice mismatch of approximately 7\%\cite{Kubota1:2011}, which can cause lattice dislocations reducing
the crystalline ordering.
An insertion of a 4~{\AA}  metallic Mg layer interfacing MgO and Mn--Ga promotes a slight
increase of the tunneling magnetoresistance ratio, which reached 23\% in Mn$_{1.63}$Ga\cite{Kubota1:2011}. The insertion is believed to
improve the crystalline quality of MgO film and
acts to diminish  the oxidation of the Mn and Ga atoms at the  interface.\cite{Kubota1:2011}
The knowledge of the actual chemical composition at the interface is essential for taking full advantage of the special spin properties
of the Mn--Ga films and reaching higher TMR ratios.

In the present work, the MgO(001)/Mn$_{1.63}$Ga(001) junction was investigated by hard
x-ray photoelectron spectroscopy (HAXPES). We found the presence of a gallium oxide layer at the interface and show that the insertion of the metallic Mg layer strongly suppresses the formation of this oxide.
HAXPES can efficiently probe burier layers \cite{Kozina:2011} and interfaces\cite{Paul:2009} because of the large escape
depth of high kinetic energy photoelectrons.
The experiments using a photon energy of h$\nu$\,=\,3.0~keV,
with a total resolution of 400~meV, were carried out in the P09 beamline of DESY (Germany); and those using h$\nu$\,=\,7.9~   keV,
with a total resolution of 250~meV, were carried out in the BL47-XU beamline of SPring-8 (Japan)\cite{Ikenaga:2013}. Fig.~1(a) depicts the
experimental geometry in both setups. The angle between the electron spectrometer and photon propagation is fixed at $90^{\circ}$
and the photoemitted electrons were collected at angle $\theta$. Samples were
measured at RT ($\sim$ 300~K). We investigated the following structures: Cr buffer(40)/Mn$_{1.63}$Ga(30)/Mg($d_{Mg}$)/MgO(2)/AlO(1)
with $d_{Mg}$= 0 (\emph{sample~1}) and 0.4 (\emph{sample~2}), where the numbers in parentheses indicate the thickness in nm.
Films were grown using an ultrahigh-vacuum magnetron sputtering system on
MgO(001) substrate and capped with 1~nm Al layer. Growth details can be found in the Ref. \onlinecite{Kubota2:2011}.

%%%%%%%%%  Fig. (1)  %%%%%%%%%%%%%%%%%%%%%%%%%
\begin{figure}[ht]
%\center{\includegraphics[width=1\columnwidth]{Fig.1vs2TWOCOL.eps}}
\center{\includegraphics[width=1\columnwidth]{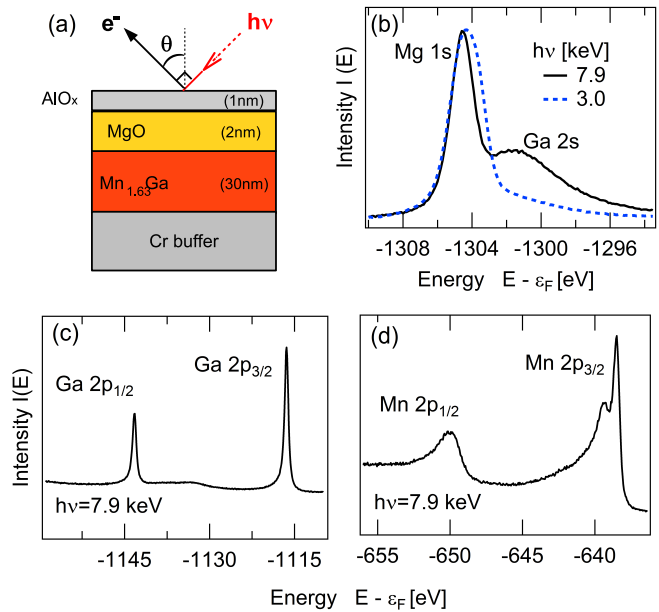}}
\vspace*{-0.6cm} \caption{\small (Color online) (a) Schematic of the experimental geometry.
The multilayered structure, representing \emph{sample~1}, was measured by the photon energy $h\nu$ and the
 photoelectrons were collected at angle $\theta$. (b)  Photoemission spectra were measured using $h\nu$\,=\,7.9~keV (solid line)
 and 3.0~keV (dashed line). The Mg 1s peak is originated from MgO and the Ga 2s peak from the buried Mn$_{1.63}$Ga
  layer. (c) Ga  and (d) Mn 2p photoemission spectra were measured using 7.9 keV photon energy. }
\label{fig:Ravg}
\end{figure}
%%%%%%%%%%%%%%%%%%%%%%%%%%%%%%%%%%%%%%%%%%%%%%%%%%%%%%%%%%%%

To understand the chemical composition near to the MgO/Mn$_{1.63}$Ga interface,
we compared the photoemission spectral shapes measured by  different photon energies.
Fig.~1(b) shows the spectra of the Mg 1s and Ga 2s measured by a photon energy of 7.9 (solid line) and 3.0~keV (dashed line).
The Ga~2s peak from the buried Mn$_{1.63}$Ga film is clearly visible in the 7.9~keV spectra, while it is only distinguishable as a
shape asymmetry of the Mg 1s peak in the 3.0~keV spectra. This is because the photon energy of 7.9~keV  produces
 photoelectrons with a large escape depth and, consequently,  positions much deeper than the interface were probed. In contrast,
 the probed depth using a photon energy of 3.0~keV extends only up to near the MgO/Mn$_{1.63}$Ga interface.
 For Ga 2s photoelectrons, the mean free path $\lambda$ through MgO and the capping layer is about 10 (3.3)~nm for a photon excitation of 7.9 (3.0)~keV, according to the TPP-2M formula\cite{Tanuma:1994}.

%Ga 2s Ek= 1696: lambda_Al= 3.5nm        lambda_MgO=3.3 nm
%Ga 2s Ek= 6636: lambda_Al=11nm         lambda_MgO=10nm

By using 7.9~keV, we measured the Ga and Mn 2p core states as shown  in Figs.~1(c) and (d), respectively. The presence of Ga oxides are commonly identified by a shoulder at high binding energy tail of the main Ga 2p$_{3/2}$ peak, as will be discussed later. In Fig. 1(c), a direct inspection in the Ga 2p$_{3/2}$ peak,
 located at 1143.4 eV, excludes any substantial amount of oxides in the bulk part of the film.
%The Ga 2p$_{1/2}$, located at 1116.5$\pm$0.1, has no particular features.
  In Fig. 1(d), the Mn 2p$_{1/2}$ state was assigned to a broad peak at about 650.0 eV, while Mn 2p$_{3/2}$ exhibited a sharp double-peak structure around 638.8 eV. The higher binding energy peak of the 2p$_{3/2}$ level originated from poorly screened states, while the lower binding energy peak was associated with well-screened final states\cite{Fujii:2011}. The relative high intensity of the well-screened peak %in comparison to the poorly-screened peak
  is not usually observed in the presence of Mn oxides, which exhibits a Mn 2p$_{3/2}$ peak located at energies above 640~eV\cite{Nelson:2000,Hishida:2013}. The Ga and Mn 2p spectra shown in Figs
  ~1(c) and (d) can therefore be safely assumed as references of the spectral shape expected for a homogeneous Mn$_{1.63}$Ga film.

%%%%%%%%%%%%%%%%%%%%%%%  Fig. (2)  %%%%%%%%%%%%%%%%%%%%%%%%%
\begin{figure}[hb]
\center{\includegraphics[width=0.85\columnwidth]{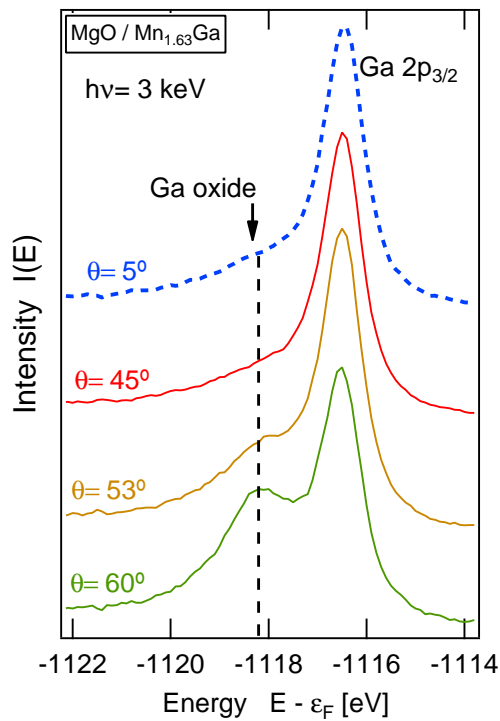}}
\vspace*{-0.0cm} \caption{\small (Color online):
Ga 2p$_{3/2}$ spectra for \emph{sample~1} measured at different photoemission angles. Curves are shown normalized by the peak height.
The arrow indicates the position of the oxide peak. Curves are vertically offset for clarity. }
\label{fig:2}
\end{figure}

%%%%%%%%%%%%%%%%%%%%%%%%%%%%%%%%%%%%%%%%%%%%%%%%%%%%%%%%%%%%

Information about the depth profile of the interface was obtained by comparing
near-normal ($\theta\sim0$) and off-normal ($\theta\gg0$) photoemission spectra using 3~keV photon energy. Electrons emitted at off-normal directions travel longer distances inside the material, for this reason they are
more subjected to scattering, resulting in an  escape depth  of approximately 3$\lambda$cos($\theta$). 
%In the setup used for experiments at 3 keV, the only possibility to perform such measurements was by changing the sample polar %angle, and consequently the photon incidence angle. Therefore the polarization conditions and relationship between peak intensities %are altered. Nevertheless we can still rely our conclusion on comparison between the two samples, measured in the same %geometry, as $\theta$ is increased.
Fig.~2 shows the Ga~2p$_{3/2}$ spectra for the \emph{sample~1} measured from $\theta=5^\circ$ to $60^\circ$.
Spectra measured at $\theta=5^\circ$ (dashed line) presented a small shoulder at higher binding energies consistent with the presence of  Ga oxides near the interface. The shoulder was due to the overlap of the main peak with the oxide peak,   which chemically shifted to higher binding energies, as reported for oxidized GaAs films\cite{Massies:1985,Paul:2009}. As $\theta$ increases, the probe depth is reduced and the contribution of the interface is enhanced. At $\theta=45^\circ$, the shoulder slightly increased. From $\theta=53^\circ$ the shoulder strongly increased until it became a distinguishable peak at $\theta=60^\circ$, which was located 1.7 eV below the main peak. The sudden onset of the Ga oxide peak between $\theta=53^\circ$ and $\theta=60^\circ$ can be explained by a sharp concentration of the oxide at the interface, which contributes to the photoemission spectra at high photoemission angles (short electron escape depth).

%%%%%%%%%%%%%%%  Fig. (3)  %%%%%%%%%%%
\begin{figure}[ht]
\center{\includegraphics[width=1\columnwidth]{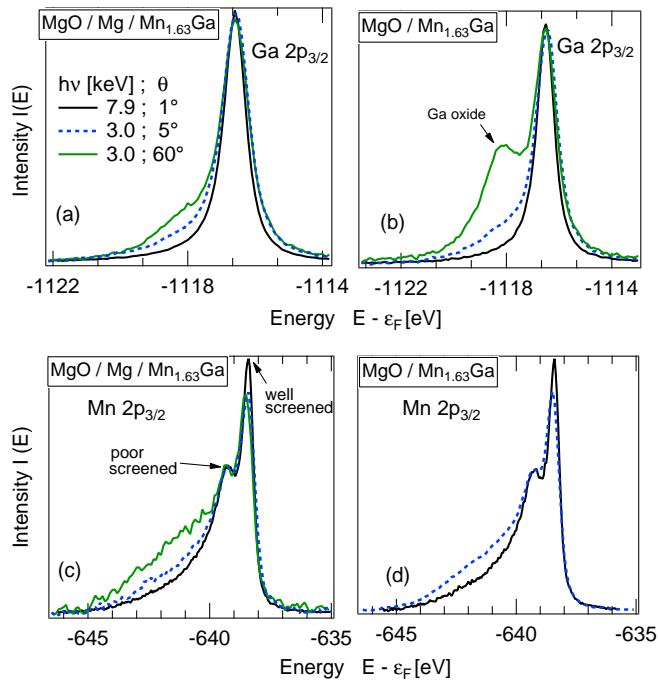}}
\vspace*{-0.5cm} \caption{\small (Color online):
Comparison between and \emph{sample~2} and \emph{sample~1}:
 Ga 2p$_{3/2}$ spectra measured with $h\nu$=7.9 and 3.0~keV for \emph{sample~2} (a) and \emph{sample~1} (b). In (a), the Ga 2p$_{3/2}$ peak  presents an increased shoulder on the high binding energies side for $h\nu$=3.0~keV (a), but the onset of the Ga oxide peak at $\theta=60^\circ$, as indicated by the arrow in (b), is not observed in (a). Mn 2p$_{3/2}$ spectra broadens towards higher-energy from $h\nu$=7.9~keV to 3.0~keV for \emph{sample~2}(c) and \emph{sample~1}(d).  In both samples, the  intensity of the well-screened peak is reduced in the measurements using $h\nu$=3.0~keV.
Shirley background was subtracted from all curves.}
\end{figure}
%%%%%%%%%%%%%%%%%%%%%%%%%%%%%%%%%%%%%%%%%%%%%%%%%%%%%%%%%%%%

In Fig.~3, we show the influence of the insertion of the Mg metallic layer by comparison of the spectra obtained for the \emph{sample~2}; (a) and (c); and \emph{sample~1}; (b) and (d). Figs.~3(a) and (b) show  the Ga 2p$_{3/2}$ spectra measured in different configurations. For both samples, it was observed an increase of the high binding energy shoulder from 7.9 to 3.0~keV photon energy in near-normal photoemission. However, in contrast with  \emph{sample~1}, \emph{sample~2} does not exhibit a distinguishable oxide peak at high photoemission angles ($\theta=60^\circ$). The absence of a clear oxide peak in the Fig.~3(a) implies that the insertion of metallic Mg layer suppressed the formation of Ga oxide at the interface.

Fig.~3(c) shows the Mn 2p$_{3/2}$ spectra for  \emph{sample~2}. From 7.9 to 3.0~keV, we observed a broadening of the spectra towards high binding energies. We also noticed a reduction of the well-screened peak with respect to the poorly screened peak. Interestingly, the broadening increased at $\theta=60^\circ$, which means that this feature is enhanced in the proximity of the MgO/Mn$_{1.63}$Ga interface. The relative intensity of the double peaks was unaltered by changing the photoemission angle. In Fig.~3(d), the Mn 2p$_{3/2}$ spectra measured with 7.9~keV photon energy for  \emph{sample~1} [shown also in Fig.~1(d)]
is compared with the measurement using 3.0~keV photon energy. The same trends observed for \emph{sample~2} were also present here.

The comparison between spectra measured at different photoemission angles was not conclusive in respect to the presence of Mn oxides.
Clearly, the local environment of Mn atoms changes from deep positions inside the film to that near the MgO interface independently on the presence of the Mg layer. The reduction of the well-screened peak intensity and broadening of the Mn 2p$_{3/2}$ spectra, when measured by the 3~keV photon energy, indicate a weakening of the hybridization of Mn valence states near the interface. Concerning the Ga oxides, their high concentration at the interface of \emph{sample~1} can be explained by the hypotheses of a Ga-rich interface. The Mn$_{3}$Ga structure is composed of periodically stacked \emph{Mn--Mn} and \emph{Ga--Mn} atomic layers. The interface with the \emph{Ga--Mn} termination plane is thermodynamically the most stable according to recent ab-initio calculations\cite{Miura:2013}, as illustrated in Fig.~4(a). Therefore,  Mn$_{1.63}$Ga might present a \emph{Ga--Mn} termination plane, which is formed mostly by Ga atoms due to the low Mn content. This would prompt the formation of Ga oxides at the interface.
The oxidation of Ga  is strongly reduced by using the metallic Mg spacer layer as depicted in Fig.~4(b).

%%%%%%%%%%%%%%%  Fig. (4)  %%%%%%%%%%%
\begin{figure}[ht]
\center{\includegraphics[width=1\columnwidth]{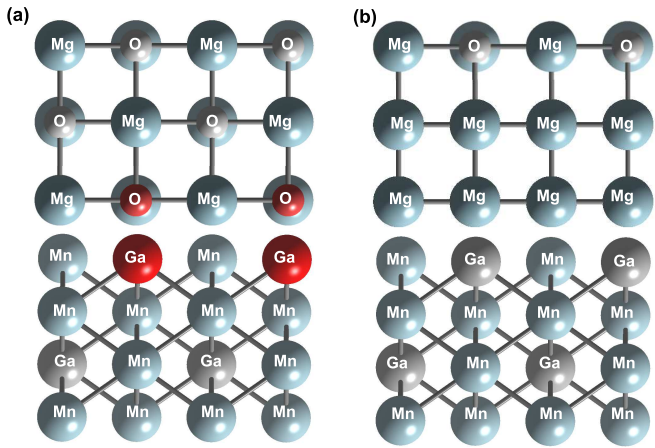}}
\vspace*{-0.4cm} \caption{\small (Color online) (a) Sketch of the Mn$_{3}$Ga/MgO interface terminations: Interface atoms are shown in red to highlight
 the predisposition for the gallium oxidation. (b) Mn$_{3}$Ga and MgO interfaced by  the metallic Mg layer, which prevents the oxidation of the Ga-rich interface.}
\end{figure}
%%%%%%%%%%%%%%%%%%%%%%%%%%%%%%%%%%%%%%%%%%%%%%%%%%%%%%%%%%%%

At this point,  the investigation of other Mn$_2$YZ systems providing better interface properties with MgO would be an interesting option.
The Mn$_3$Ge is one feasible candidate\cite{MizukamiMnGe:2013}, since ab-initio calculations predicted high TMR ratios independently on the interface
termination\cite{Miura:2013}.  Alternatively, the engineering of compounds using main group elements Z with larger atomic radii, as Sb and Sn,
would favor better lattice matching with MgO. The optional configuration should also take into account the reactivity of the elements
with  oxygen. The combination of these properties will be the key for reaching ultra-high TMR ratios in Heusler-base MTJ,
consisting in a challenge for both theoreticians and experimentalists.

In summary, we investigated the MgO/Mn$_{1.63}$Ga junction grown by sputtering deposition and the influence of the insertion of a thin metallic Mg layer at the interface. Depth-resolved chemical information could be obtained by changing the photon energy and photoemission angle in the HAXPES experiment. The local environment of Mn atoms strongly changes near the MgO interface. The experiment points to a reduction of the hybridization for shallow Mn atoms. This behavior seems not to be affected by the insertion of the metallic Mg layer. The angular dependence of photoelectrons provides evidences for the formation of Ga oxide concentrated at MgO/Mn$_{1.63}$Ga interface.
The  formation of Ga oxide is suppressed by the insertion of the metallic Mg layer.  We ascribe the presence of Ga oxide to a Ga-rich interface due to the low Mn content in the Mn$_{1.63}$Ga film. This information will stimulate the investigation of new growth methodologies and engineering of MTJs.

\section*{Acknowledgments}
Financial support by Deutsche Forschungsgemeinschaft and the Strategic
International Cooperative Program of JST (DFG-JST) (Project P1.3-A of research
unit FOR 1464: {\it ASPIMATT}) is gratefully acknowledged. HAXPES experiments
were performed at BL47XU of SPring-8 with approval of JASRI (Proposal
No.~2012B0043) and at P09 of DESY (Project No.~I-20120684).

\bibliography{Myrefbib2}

\end{document}